# An Integrated Impact Indicator (*I3*):

# A New Definition of "Impact" with Policy Relevance

Caroline S. Wagner[a] and Loet Leydesdorff[b]

**Abstract**

Allocation of research funding, as well as promotion and tenure decisions, are increasingly made using indicators and impact factors drawn from citations to published work. A debate among scientometricians about proper normalization of citation counts has resolved with the creation of an Integrated Impact Indicator *(I3)* that solves a number of problems found among previously used indicators. The *I3* applies non-parametric statistics using percentiles, allowing highly-cited papers to be weighted more than less-cited ones. It further allows unbundling of venues (i.e., journals or databases) at the article level. Measures at the article level can be re-aggregated in terms of units of evaluation. At the venue level, the *I3* creates a properly weighted alternative to the journal impact factor. *I3* has the added advantage of enabling and quantifying classifications such as the six percentile rank classes used by the National Science Board's *Science & Engineering Indicators*.

[a] John Glenn School of Public Affairs, The Ohio State University, Columbus, OH 43210 USA
wagner.911@osu.edu; www.careolineswagner.org
[b] Amsterdam School of Communications Research (ASCoR), Kloveniersburgwal 48, 1012 CX Amsterdam, The Netherlands; loet@leydesdorff.net; http://www.leydesdorff.net.

**Introduction**

A refereed exchange among scientometricians about appropriate normalization (Gingras & Larivière, 2011) has resulted in the creation of a refined indicator that solves a number of problems that arise when assessing the citation impact of scientific articles and venues. Citation and publication distributions are well-known to be heavily skewed (Seglen, 1992, 1997). Following the prime example of the impact factors, however, scientometric indicators have been based on using averages. The impact factor, for example, was defined by Garfield (1972; cf. Sher & Garfield, 1955) as the number of citations in a given year to the citable items in a venue during the two preceding years. Journals are then compared in terms of central-tendency statistics.

Using percentiles (deciles, quartiles, etc.) one is able to compare skewed distributions. It is possible to organize percentile rank classes such as the top-1%, top-5%, etc., the method used for more than a decade in the *Science & Engineering Indicators* of the U.S. National Science Board (2012; Bornmann & Mutz, 2011). Non-parametric statistics make it possible to test whether the percentile scores are above or below expectation, and also to test whether differences among two units (journals, departments) are statistically significant (Bornmann *et al.*, 2012; Leydesdorff & Bornmann, 2012). The percentage of top-1% or top-10% most-highly cited papers, for example, can also be considered as an Excellence Indicator (Tijssen *et al.*, 2006; Waltman *et al.*, 2012; cf. SCImago Institutions Rankings at http://www.scimagoir.com/pdf/sir_2011_world_report.pdf.)

The *Integrated Impact Indicator* (*I3*) provides a framework for organizing these percentile-based indicators.[1] *I3* can formally be written as follows: $I3 = \sum_i x_i * n(x_i)$, in which $x_i$ denotes the percentile (rank) value $i$, and $n$ the number of papers with this value. The ordering in terms of six percentile rank classes (*PR6*) such as the ones used by the NSF or in terms of an excellence indicator follow from *I3* as aggregations. The top-10% most highly-cited papers—used increasingly as an Excellence Indicator (*EI*)—can be considered as a special case of *I3* in which only two percentile rank classes are distinguished and weighted with zero and one, respectively (Rousseau, 2012).

This article provides examples of the application of *I3* and *PR6* at the researcher and venue levels. Assuming that a decision maker or research manager wishes to use publications and citations as output indicators (Donovan, 2011; cf. Bornmann & Daniel, 2008), the changes in measurement and ranking come down to definitions and mathematical principles. It has long been known that publication rankings require normalization because of differences in publication numbers and citation practices across fields of science (Garfield, 1979). Practitioners in all fields, however, acknowledge one another's work by citing influential papers. As a single paper accrues more citations, it is assumed to be higher in quality and thus of higher impact. (There are notable exceptions to this rule, such as the Fleishman-Pons claim for nuclear fusion at room temperature, but negative citations are the exception in science (Bornmann and Daniel, 2008).)

---

[1] The percentiles can be considered as a continuous random variable (quantiles; Rousseau, 2012).

Just as publication frequencies differ across fields, so do citing norms and patterns. To account for these differences, it is standard practice to normalize by field or at the venue level creating an average of relative citations. The average of citations per publication (*c/p*) has the obvious *disadvantage* that the total number of publications is the denominator, greatly watering down the impact factor for the few highly-cited papers. (For example, when one adds to a principal investigator (PI) the less-cited papers of other members of his team, the average impact will go down because of the larger *N* in the denominator.) It is nearly always the case that citation distributions are skewed, with a few papers garnering many citations and most papers receiving one or none.

But what is the appropriate procedure if two PIs have different publication and citation profiles? Can two papers in the $39^{th}$ percentile be considered as equivalent to one in the $78^{th}$ or is a non-linearity involved (as in the case of the six percentile rank classes)? In Figure 1, we compare the citation curves of two principal investigators of the Academic Medical Center of the University of Amsterdam. In this academic hospital the *c/p*-ratios are used in a model to allocate funding, raising the stakes for methods of assessing impact and inciting the researchers to question the exactness of the evaluation (Opthof & Leydesdorff, 2010). The *Integrated Impact Indicator* quantifies the skewed citation curves by normalizing the documents first in terms of percentiles. The question of the normative scheme used for the evaluation can then be considered as the specification of an aggregation rule for the binning and weighting of these scores.

**Impact at the Level of the Individual Researcher**

Figure 1 shows the outputs of two PIs: PI 1 has 1,623 citations from 23 papers and PI 2 has 1,578 citations from 65 papers. For analytical reasons, integration of the surfaces underneath the citation curves in the left-hand figure provides the total numbers of citations. Whereas the average *c/p* ratio of PI 1 is 1,632/23 = 70.96 against 1,578/65 = 24.28 for PI 2, the total numbers of citations are not so different. However, the alternative of using the total number of citations without normalization does not yet qualify highly-cited papers as different from less highly-cited ones.

As the right-hand figure shows, normalization of each paper in terms of the percentile ranks obtained in the different journals in which they are respectively published (after proper control for the same publication year and document type) changes the picture. The integration of the normalized citation distributions provides the *Integrated Impact Indicator I3* and shows that PI 2 has a higher overall impact (Table 1).

**Table 1**: PI 1 and PI 2 compared in terms of the six percentile classes used by NSB

| Percentile rank | Weight of rank ($x_i$) | PI 1 ($n_i$) | PI 1 ($n_i * x_i$) | PI 2 ($n_i$) | PI 2 ($n_i * x_i$) |
|---|---|---|---|---|---|
| top-1% | 6 | 3 | 3x6 = 18 | 0 | 0x6 = 0 |
| 95-99% | 5 | 3 | 3x5 = 15 | 5 | 5x5 = 25 |
| 90-95% | 4 | 1 | 1x4 = 4 | 1 | 1x4 = 4 |
| 75-90% | 3 | 3 | 3x3 = 9 | 10 | 10x3 = 30 |
| 50-75% | 2 | 6 | 6x2 = 12 | 14 | 14x2 = 28 |
| 0-50% | 1 | 7 | 7x1 = 7 | 35 | 35x1 = 35 |
| Total | | 23 | $\sum_i^6 x_i n_i = 65$ | 65 | $\sum_i^6 x_i n_i = 122$ |

The difference between the scores for these two PIs is statistically significant (Leydesdorff, Bornmann, Mutz, and Opthof, 2011). Normalization in terms of percentiles greatly improves comparisons across articles at the level of individual researchers and research groups. Using this normalization, for example, a group of researchers has a citation impact equal to the sum of the impacts of the group members. Furthermore, an impact measure, in our opinion, should correlate strongly with both the number of publications and citations. When one averages and thus divides the number of citations by the number of publications, one can expect to lose the correlations with each of these two indicators in the numerator and denominator, respectively (Leydesdorff, 2009).

In addition to using hundred percentiles (as a continuous random variable), the six classes (top-1%, top-5%, top-10%, top-25%, top-50%, and bottom-50%) can be obtained by simple aggregation of the weighted rank classes as provided, for example, in the National Science Board's *Science and Engineering Indicators* (2010, Appendix Table 5-43). However, it is also possible to use deciles or quartiles once the percentile values are known at the article level. Thus, the choice of a normative framework for the evaluation is not pre-determined by the analysis.

**Impact at the Venue Level**

Scientometricians often normalize at the venue level (i.e., journals or sets of journals) using the field classification systems in the *Science Citation Index* and *Scopus*. Publishers regularly advertise their "journal impact factor" (JIF) to improve the quality of submissions. Impact factors, however, can be considered as two-year averages over skewed distribution and therefore one can expect problems of unfair evaluations similar to the problems with *c/p* ratios of individual researchers and research groups.

In Figure 2, the 48 journals classified in the *Science Citation Index* as "multidisiciplinary" are used as the reference set to compare the three leading journals in this category (*Science*, *Nature*, and *Proceedings of the National Academy of Sciences PNAS*). The left-hand panel shows the raw citation curves for 2009 of the articles in the three journals during 2007 and 2008; the *c/p* values are then by definition equal to the JIFs 2009. The visual shows *Science* and *Nature* competing for first place. Then—using the same data—the right-hand panel shows results with the normalization in terms of percentiles. *Science* and *Nature* still have nearly identical curves, but *Proceedings of the National Academy of Sciences PNAS* stands out as having significantly higher impact. (The values for the IFs, *I3* and six percentile ranks (*PR6*) are summarized in Table 2.)

**Table 2**: "Multidisiciplinary" journals with highest values for *I3* and six percentile ranks (6PR) compared in rankings (between brackets) with journal impact factors and total citations. Source: Leydesdorff & Bornmann, 2011.

| Journal | N of papers (a) | N of citations (b) | %I3 (c) | % PR6 (d) | JIF 2009 (e) |
|---|---|---|---|---|---|
| *Proc Nat Acad Sci USA* | 7,058 | 178,137 | 43.29 [1] + | 33.64 [1] + | 9.432 [3] |
| *Nature* | 2,285 | 150,718 | 16.31 [2] + | 16.46 [2] + | 34.480 [1] |
| *Science* | 2,253 | 126,230 | 15.68 [3] + | 15.27 [3] + | 29.747 [2] |
| *Ann NY Acad Sci* | 1,996 | 14,284 | 9.33 [4] + | 8.29 [4] | 2.670 [5] |
| *Curr Sci* | 1,271 | 1,551 | 2.33 [5] - | 3.40 [5] - | 0.782 [22] |
| *Chin Sci Bull* | 1,115 | 2,239 | 2.11 [6] - | 2.55 [6] - | 0.898 [20] |

+ above expectation at *p*< 0.01; - below expectation at *p*< 0.01 (using the *z*-test).

To contrast the results obtained when using the new *I3* and the JIFs, Table 2 compares six prestigious journals which target a broadly-interested readership and with highest citation impacts using the percentile ranking. Since absolute values of *I3* and *PR6* are based on summations, we use their relative values as percentages for clarity. When using these new indicators (columns c and d), *PNAS* has a much higher impact than would be derived by using an average-based JIF (shown in column e). Indeed, the finding of much higher impact factors for *Science* and *Nature* (column e) are an artifact of the smaller numbers of publications (column a) in these two journals rather than higher citation counts at the top end. All three journals, however, have an impact significantly above expectation ($p < 0.01$).

In contrast, consider the next three journals in Table Two. The *Annals of the New York Academy of Sciences* follows at the fourth position in terms of *I3*, but if the six percentile ranks of the NSB are used, *Annals* no longer scores significantly above expectation. *PR6* gives more weight to top-cited papers than *I3*. The two Asian journals in the category—the *Chinese Science Bulletin* and the Indian journal *Current Science*—are ranked at the fifth and sixth positions among this group of 48 "multidisciplinary" journals, while they were ranked much lower—20$^{th}$ and 22$^{nd}$, respectively—using JIFs, as can be seen in column (e). The citation rates of these two journals, however, are still below expectation.

**Table 3**: Rank-order correlations (Spearman's *ρ*; upper triangle) and Pearson correlations *r* (lower triangle) for the 48 journals attributed to the WoS Subject Category "multidisciplinary sciences" in 2009. Source: Leydesdorff & Bornmann, 2011, p. 2142).

| Indicator | IF-2009 | I3 | PR6 | Number of publications | Total citations |
|---|---|---|---|---|---|
| IF-2009 |  | .798 ** | .517 ** | .479 ** | .840 ** |
| I3 | .590 ** |  | .854 ** | .829 ** | .986 ** |
| PR6 | .660 ** | .987 ** |  | .996 ** | .801 ** |
| N of publications | .492 ** | .953 ** | .967 ** |  | .772 ** |
| Total citations | .841 ** | .922 ** | .945 ** | .839 ** |  |

Note: ** Correlation is significant at the 0.01 level (2-tailed); * Correlation is significant at the 0.05 level (2-tailed).

If the comparison is make among JIFs with *I3* and *PR6* values for the full set of 48 journals in this set (with *N* of documents is equal to 24,494), the Pearson correlations are .590 and .660 ($p < 0.01$), respectively. As can be expected *I3* and *PR6* are highly correlated among them ($r = .987$), as they are both referencing citation and publication rates. All these correlations with productivity and impact are larger than .9. However, JIF correlates .49 with the number of publications and .84 with the number of citations. In other words, the division by the number of publications makes *average* impact different from impact, and this change in the semantics matters in evaluation scenarios.

**An Essential Change: impacts add up instead of averaging out**

Before this change in the definition of impact, it was common to use two conventions for normalization: (1) normalization in terms of fields of science, and (2) comparison of a paper's or journal's citation rate to the world average. Both of these conventions raise problems when assessing impact.

The sources of error in the first practice—normalizing in terms of a field—come from using journals such as *Nuclear Physics B* or *Cell* as the units for normalization: a number of studies have demonstrated that specialist journals do not necessarily represent a single discipline (Boyack & Klavans, 2011; Pudovkin & Garfield, 2002; Rafols & Leydesdorff, 2009). Therefore, even if the *Science Citation Index* and *Scopus* refined the journal classifications, it is not clear that this would solve the problem of field delineation or improve the quality of rankings (Leydesdorff, 2006).

The second convention—comparison to the world average—was defined early in the scientometric enterprise by Schubert & Braun (1986) who proposed to compare the mean observed citation rate (*MOCR*) within a database (representing, for example, a field of science) with the corresponding mean *expected* citation rate (*MECR*) as the average citation rates of papers of the same datatype (reviews, articles, or letters) and publication year. The Relative Citation Rate (= *MOCR/MECR*) is thus normalized as the world average.

The combination of these measures caused the problems: The division of two means results in mathematical inconsistency because the order of operations says that one should divide first and then sum, not sum and then divide the averages. As this error became clear (Opthof & Leydesdorff , 2010; cf. Lundberg, 2007), the renowned Leiden Centre for Science and Technology Studies (CWTS) changed its main ("crown") indicator (van Raan *et al*., 2010). CWTS called this "new crown indicator" the Mean Normalized Citation Score or *MNCS*. One advantage of the new indicator is that the mean is a statistic with a standard deviation, and consequently a standard error of the measurement can be defined and can be published as an error bar in relevant assessments. Waltman *et al*. (2011) showed that this new indicator is mathematically consistent.

To further refine the indicator for broad application, Leydesdorff & Bornmann (2011) elaborated this approach by defining the *Integrated Impact Indicator* (*I3*). *I3* leaves the parametric domain of working with averages behind and moves to non-parametric statistics using percentiles. Rousseau (2012) discussed the mathematical properties of the new indicator in more detail. Unlike the *h*-index, the tails of the distribution are also weighted in *I3*, and standard statistics (as available, for example, in SPSS) can be used.

When creating the ISI-impact factor in the early days of the *Science Citation Index*, Eugene Garfield (e.g., 1972) deliberately chose to normalize by dividing by *N* in order to prevent the larger journals from overshadowing the smaller. Bensman (2007) found "Total Citations" to be more closely correlated than the JIFs with subjective appreciation of faculty. Even so, Total Citations and Impact Factors were crude (first-generation?) measures. The percentile approach allows a user both to account for the skewed distribution of citations and appreciate differences among highly-cited papers and less highly-cited ones. As said, the user can then aggregate the percentiles in a normative evaluation scheme (e.g., quartiles or the six classes of the NSF).

**Normative implications and policy relevance**

Research funds are often allocated based upon these measures. Fields of science, institutions, and nations are increasingly ranked based upon the recognition bestowed by citation counts. Policy decisions about spending the incremental research dollar, euro, or yen often rest upon the excellence of research units in terms of citation counts and impacts (Hicks, 2012). Thus, these distinctions are important to users across a wider spectrum of research evaluation and policy.

Indicators clearly matter to the individual researchers, as discussed in the comparison of PI 1 and PI 2 above; research units and nations are also served by the improvement offered by *I3*. Using a refined indicator can improve the efficiency of research spending by increasing the likelihood that the most relevant or appropriate researcher or research unit receives funding. In times of difficult budget choices, it is even more important to ensure the accuracy of the basic measures of the research system and its components—the benefit offered by using the Integrated Impact Indicator.

Additional information and free software for the automatic analysis of document sets in terms of the percentile values can be found online at http://www.leydesdorff.net/software/i3/index.htm.

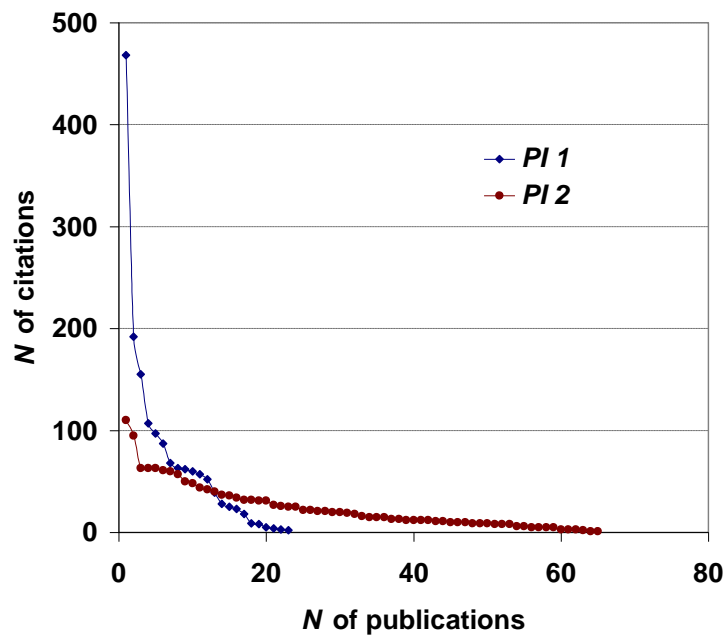 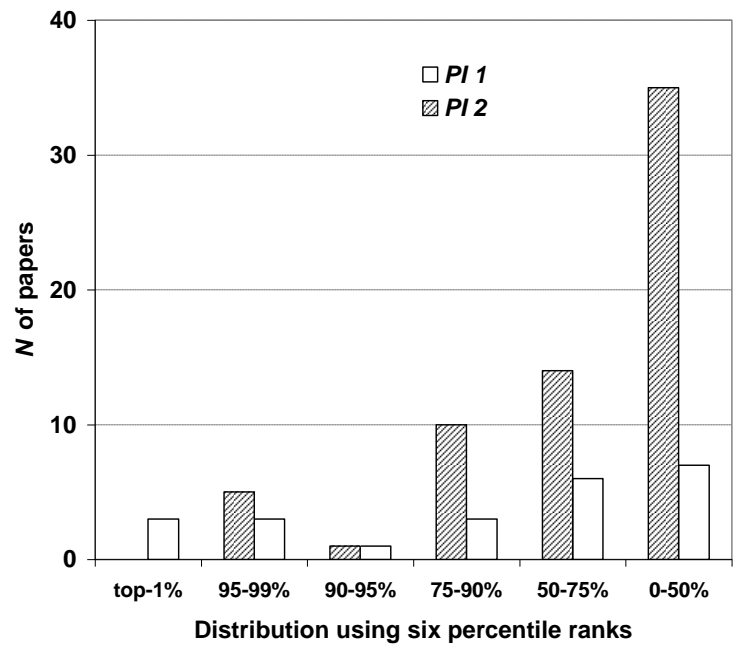

**Figure 1**: Citation curves and percentile ranks for 23 publications of *PI 1* and 65 publications of *PI 2*, respectively.

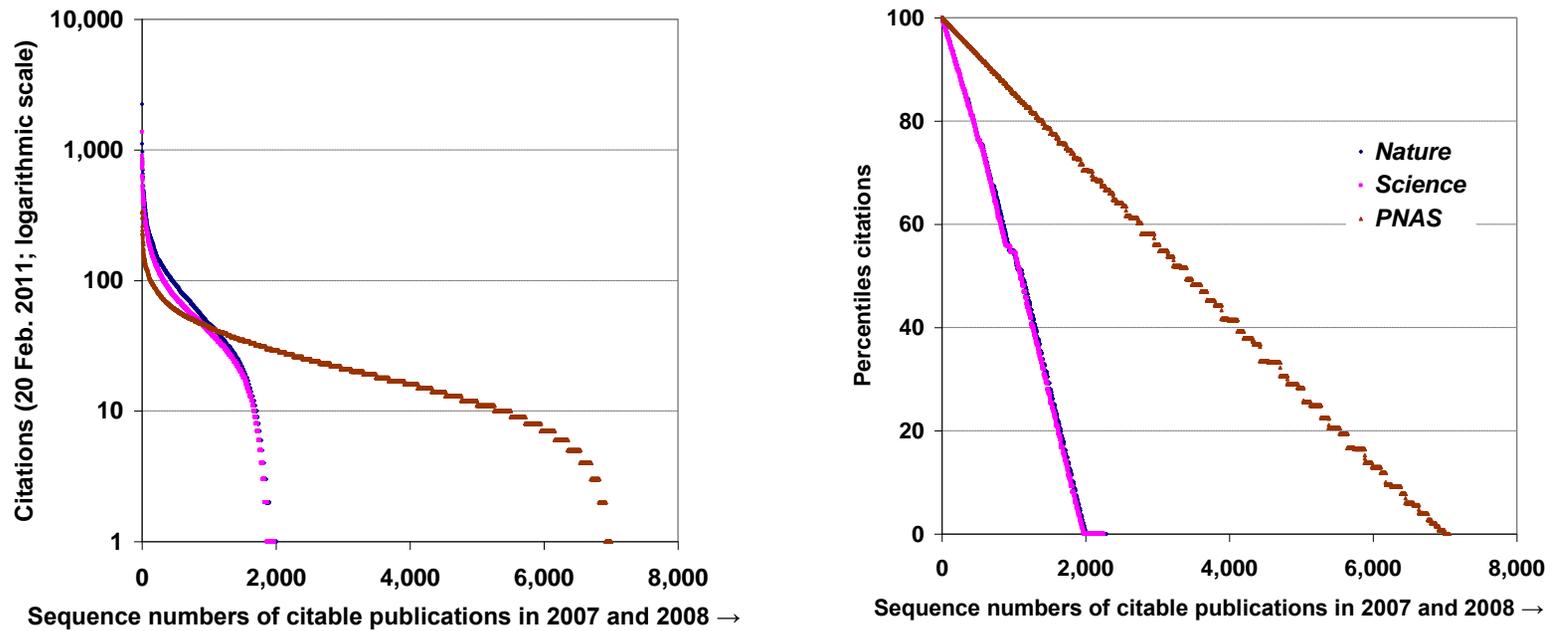

**Figure 2**: Citation rates and percentiles for *Nature* (♦), *Science* (■), and *PNAS* (▲), respectively; using 48 "multidisciplinary" journals in the *Science Citation Index* as the reference set. (Source: Leydesdorff & Bornmann, 2011, p. 2141.)